\documentclass[twocolumn,aps,showpacs,amsmath,superscriptaddress]{revtex4-1} 
\usepackage{bm}
\usepackage{amsmath}
\usepackage{amssymb}
\usepackage{times}
\usepackage[usenames,dvipsnames]{color}
\usepackage{graphicx}
\usepackage{dcolumn}
\usepackage{simplewick}
\usepackage{hyperref}
\hypersetup{
  colorlinks=false,
  colorlinks=true,
  citecolor=blue,
  linkcolor=blue,
  urlcolor=blue}

\begin{document}

\title{Electric dipole polarizability of group-IIIA ions using PRCC: Large 
       correlation effects from nonlinear terms}

\author{Ravi Kumar}
\affiliation{Department of Physics, Indian Institute of Technology, 
             Hauz Khas, New Delhi 110016, India}

\author{S. Chattopadhyay}
\affiliation{Department of Physics and Astronomy, Aarhus University,
             DK-8000, Aarhus C
             }

\author{B. K. Mani}
\affiliation{Department of Physics, Indian Institute of Technology, 
             Hauz Khas, New Delhi 110016, India}

\author{D. Angom}
\affiliation{Physical Research Laboratory,
             Ahmedabad - 380009, Gujarat,
             India}

\begin{abstract}

We compute the ground-state electric dipole polarizability of group-IIIA ions 
using the perturbed relativistic coupled-cluster (PRCC) theory. To account for
the relativistic effects and QED corrections, we use the Dirac-Coulomb-Breit 
Hamiltonian with the corrections from the Uehling potential and the self-energy. 
The effects of triple excitations are considered perturbatively in the PRCC. 
Our PRCC results for $\alpha$ are good in agreement with the previous theoretical
results for all the ions. From our computations we find that the nonlinear terms in 
PRCC have significant contributions and must be included to obtain the accurate 
value of $\alpha$ for group-IIIA ions. For the correction from the Breit 
interaction, we find that it is largest for Al$^+$ and decreases as we go 
towards the heavier ions. The corrections from the vacuum polarization and
the self-energy increase from lighter to heavier ions.

\end{abstract}



\maketitle

\section{Introduction}

  The electric dipole polarizability ($\alpha$) of atoms or ions is a 
measure of the interaction with an external electromagnetic field 
\cite{bonin-97}. It is a key parameter, and plays an important role in probing 
fundamental as well as technologically relevant properties of atoms and ions. 
Some current and potential implications of $\alpha$ in atomic systems include 
discrete symmetry violations in atomic systems 
\cite{khriplovich-91, griffith-09}, optical atomic 
clocks \cite{udem-02,diddams-04}, condensates of dilute atomic 
gases \cite{anderson-95,bradley-95,davis-95}, and the search  for the
variation in the fundamental constants \cite{karshenboim-10}.

The recent advances in development of new and improved frequency and time
standards in optical domain has elevated the interest in electric dipole
polarizability of atoms and ions. One of the important reasons for this is that the 
$\alpha$ is essential to calculate the blackbody radiation (BBR) shift in 
atomic transition frequency due to ac Stark effect. The BBR shift is one of the 
dominant environment induced shifts in atomic transition frequency, and 
contributes to the inaccuracy of atomic clocks. Here, it is to be emphasized 
that the group-IIIA ions are the promising candidates for accurate 
optical atomic clocks as they are expected to have low fractional frequency 
errors \cite{chou-10,safronova-11b,zuhrianda-12}. Despite this important
prospect associated with the group-IIIA ions, the ground state polarizability 
of these ions have not been studied in detail. For example, except for Al$^+$, 
very little data is available from the previous theoretical calculations. 
This, perhaps, can be attributed to the complex nature of the correlation 
effects in these divalent ions. 

  It can thus be surmised that there is a research gap on the dipole 
polarizability for the group-IIIA ions. But, considering the 
experimental developments there are compelling reasons to address this 
research gap. That is the aim of this work. For this we employ the perturbed 
relativistic coupled-cluster (PRCC) theory and compute the ground state 
$\alpha$ of the group-IIIA ions and examine the trends in the correlation 
effects in detail. More precisely, our aim is to: compute accurate value of 
$\alpha$ for B$^+$, Al$^+$, Ga$^+$, In$^+$ and Tl$^+$ ions using the PRCC 
theory; examine in the detail the contributions from the nonlinear terms in 
PRCC theory; do a comparative study with the trends observed in the 
other closed-shell atoms and ions 
\cite{chattopadhyay-12a, chattopadhyay-12b, chattopadhyay-13a,
chattopadhyay-13b,chattopadhyay-14,chattopadhyay-15}; and 
examine in detail the contributions to $\alpha$ from the Breit 
interaction, the vacuum polarization and the self-energy corrections,
and compare with the contributions in other closed-shell atoms and ions. 

 The PRCC theory is an appropriate many-body theory to account for the 
correlation effects arising from the external perturbation. It has been 
used to compute accurate $\alpha$ for several atoms and ions in a series
of our previous works \cite{chattopadhyay-12a,chattopadhyay-12b,
chattopadhyay-13a, chattopadhyay-13b,chattopadhyay-14,chattopadhyay-15}. The
essence of PRCC is that it is a relativistic coupled-cluster (RCC) theory 
\cite{pal-07, mani-09,nataraj-11} with an additional set of cluster operators. 
The latter account for the effects of an internal or external perturbation
Hamiltonian. The amplitudes of these cluster operators are obtained by solving 
a new set of coupled nonlinear equations, this is in addition to the RCC 
cluster amplitude equations. The added advantage of PRCC is that it does 
not employ the sum-over-state \cite{safronova-99,derevianko-99} approach 
to incorporate the effects of a perturbation. The summation over all the 
possible intermediate states is subsumed in the perturbed cluster operators. 
In our previous works we have also demonstrated and verified the 
implementations of Breit interaction \cite{chattopadhyay-12b}, vacuum 
polarization \cite{chattopadhyay-13b}, and triple excitation in unperturbed 
\cite{chattopadhyay-14} and perturbed \cite{chattopadhyay-15} cluster 
operators. In the literature, there are other many-body theories which have
been used to compute $\alpha$ to good accuracy for a variety of atomic 
systems. A recent review by Mitroy and collaborators \cite{mitroy-10} 
provides a detailed overview of these many-body theories and their applications. 
  The remaining part of the paper is organized as follows. In Section II we 
provide an overview of the RCC and PRCC theories. In Section III 
we provide the calculational details where we discuss about the basis 
functions, nuclear potential, etc. used in the present work. The results 
obtained from our computations are analyzed and discussed in the Section IV. 
Unless stated otherwise, all the results and equations presented in this 
paper are in atomic units ( $\hbar=m_e=e=1/4\pi\epsilon_0=1$).


\section{Theoretical Methods}

  We use the Dirac-Coulomb-Breit no-virtual-pair Hamiltonian, $H^{\rm DCB}$,
to incorporate the relativistic effects in high-Z atoms. It provides a good 
description of the structure and properties of heavier atoms and ions.  
For an $N$-electron atom or ion
\begin{eqnarray}
   H^{\rm DCB} & = & \Lambda_{++}\sum_{i=1}^N \left [c\bm{\alpha}_i \cdot 
        \mathbf{p}_i + (\beta_i -1)c^2 - V_{N}(r_i) \right ] 
                       \nonumber \\
    & & + \sum_{i<j}\left [ \frac{1}{r_{ij}}  + g^{\rm B}(r_{ij}) \right ]
        \Lambda_{++},
  \label{ham_dcb}
\end{eqnarray}
where $\bm{\alpha}$ and $\beta$ are the Dirac matrices, $\Lambda_{++}$ is 
an operator which project out the negative continuum states and 
$V_{N}(r_{i})$ is the nuclear potential. Projecting the Hamiltonian with 
$\Lambda_{++}$ ensures that the Hamiltonian is bounded from below by 
neglecting the negative-energy continuum states from the calculations. In
the present work, this is implemented by selecting only the positive energy
states from the finite size basis set. The last two terms, $1/r_{ij} $ and 
$g^{\rm B}(r_{ij})$  are the Coulomb and Breit interactions, respectively. The 
Breit interaction, which represents the inter-electron magnetic interactions, 
is 
\begin{equation}
  g^{\rm B}(r_{12})= -\frac{1}{2r_{12}} \left [ \bm{\alpha}_1\cdot\bm{\alpha}_2
               + \frac{(\bm{\alpha_1}\cdot \mathbf{r}_{12})
               (\bm{\alpha_2}\cdot\mathbf{r}_{12})}{r_{12}^2}\right].
\end{equation}
The Hamiltonian $H^{\rm DCB}$ satisfies the eigen-value equation
\begin{equation}
   H^{\rm DCB}|\Psi_{i}\rangle = E_{i}|\Psi_{i}\rangle , 
  \label{si_eqn}
\end{equation}
where, $|\Psi_{i}\rangle$ is the exact atomic state and $E_i$ is the 
corresponding exact energy. 

In the presence of external perturbations, the Hamiltonian $H^{\rm DCB}$ is 
modified with the addition of the perturbation interaction terms. For example, 
the total Hamiltonian in presence of an external electric field 
$\mathbf{E}_{\rm ext}$ is
\begin{equation}
   H_{\rm Tot} =  H^{\rm DCB} + \lambda H_{\rm int}, 
  \label{ham_tot}
\end{equation}
where $H_{\rm int}=-\mathbf{D}\cdot\mathbf{E}_{\rm ext}$ is the interaction 
Hamiltonian, arising from the interaction between the induced electric dipole
moment $\mathbf{D}$ of the atom and the external electric field 
$\mathbf{E}_{\rm ext}$. And, $\lambda$ is a perturbation parameter. The 
modified Hamiltonian satisfies the eigen-value equation
\begin{equation}
   H_{\rm Tot}|\tilde \Psi_{i}\rangle = \tilde E_{i}|\tilde \Psi_{i}\rangle, 
  \label{psi_eqn}
\end{equation}
here, $|\tilde \Psi_{i}\rangle$ and $\tilde E_{i}$ represent the perturbed 
atomic state and the corresponding perturbed eigen energy, respectively.


 To compute $|\Psi_i\rangle$  and $|\tilde \Psi_{i}\rangle$ we use 
RCC \cite{mani-09} and PRCC \cite{chattopadhyay-12a, chattopadhyay-12b,
chattopadhyay-13a, chattopadhyay-13b, chattopadhyay-14, chattopadhyay-15} theories, 
respectively. In the RCC theory the  ground state atomic wavefunction of a 
closed-shell atom is
\begin{equation}
|\Psi_0\rangle = e^{ T^{(0)}}|\Phi_0\rangle,
 \label{psi}
\end{equation}
where $|\Phi_0\rangle$ is the reference state wave-function and $T^{(0)}$ is 
the cluster operator. The perturbed ground state, based on the PRCC theory, is 
\begin{equation}
 |\tilde{\Psi}_0\rangle = e^{T^{(0)} + \lambda \mathbf{T}^{(1)}\cdot\mathbf{E}} 
 |\Phi_0\rangle = e^{T^{(0)}}\left [ 1 + \lambda \mathbf{T^{(1)}\cdot 
 \mathbf{E}} \right ] |\Phi_0\rangle . \;\;\;\;
 \label{psi-tilde}
\end{equation}
For an $N$-electron closed-shell atom $T^{(0)} = \sum_{i=1}^N T_{i}^{(0)}$ and 
${\mathbf T}^{(1)} = \sum_{i=1}^N {\mathbf T}_{i}^{(1)}$, where $i$ is the 
order of excitation. An approximation, which captures most of the correlation
effects, is the coupled-cluster single and double (CCSD) approximation 
\cite{purvis-82}. With this approximation
\begin{subequations}
\begin{eqnarray}
T^{(0)} &=& T_{1}^{(0)} + T_{2}^{(0)}, \\ 
{\mathbf T}^{(1)} &=& {\mathbf T}_{1}^{(1)} + {\mathbf T}_{2}^{(1)}.
\end{eqnarray}
\end{subequations}
The cluster operators in second quantized notations are
\begin{subequations}
\begin{eqnarray}
  T_1^{(0)} &= &\sum_{a,p} t_a^p {{a}_p^\dagger} a_a , 
                \label{t1_def}         \\
  T_2^{(0)} &= &\frac{1}{4}\sum_{a,b,p,q}
                t_{ab}^{pq} {{a}_p^\dagger}{{a}_q^\dagger}a_b a_a , 
                \label{t2_def}         \\
  \mathbf{T}_1^{(1)} & = & \sum_{a,p} \tau_a^p \mathbf{C}_1 (\hat{r})
                       a_{p}^{\dagger}a_{a},
                \label{pt1_def}         \\
  \mathbf{T}_2^{(1)} & = & \frac{1}{4}\sum_{a,b,p,q} \sum_{l,k} 
                    \tau_{ab}^{pq}(l,k) \{ \mathbf{C}_l(\hat{r}_1) 
                    \mathbf{C}_k(\hat{r}_2)\}^{1}
                   a_{p}^{\dagger}a_{q}^{\dagger}a_{b}a_{a}, \;\;\;\;\;\;\;\;
                \label{pt2_def}  
\end{eqnarray}
\end{subequations}
where $t_{\ldots}^{\ldots}$ and $\tau_{\ldots}^{\ldots}$ are the cluster 
amplitudes, $a_i^{\dagger}$ ($a_i$) are single particle creation (annihilation)
operators and $abc\ldots$ ($pqr\ldots$) represent core (virtual) 
single-particle states or orbitals. Here, we have used $\mathbf{C}$-tensors to 
represent the perturbed cluster amplitudes to incorporate the rank of 
$\mathbf{D}$ in the perturbation Hamiltonian. Besides this modification, 
$\mathbf{T}_2^{(1)}$ are also constrained by the parity and triangular
conditions  \cite{chattopadhyay-12b}: $(-1)^{(l_a + l_p)} = (-1)^{(l_b + l_q)}$;
$|j_a - j_p| \le l \le (j_a + j_p)$, $|j_b - j_q| \le k \le (j_b + j_q)$, 
and $|l - k| \le 1 \le (l + k)$. Where, $l(j)$ represents the orbital(total)
angular momentum of the single-electron state. 

   The unperturbed cluster operators $T^{(0)}$ used in equation (\ref{psi}) 
are obtained by solving the coupled nonlinear equations
\begin{widetext}
\begin{subequations}
\begin{eqnarray}
  \langle\Phi^p_a| H_{\rm N} +  \left [ H_N, T^{(0)} \right ] 
   + \frac{1}{2!} \left [\left [ H_N, T^{(0)}\right], T^{(0)}\right]  
   + \frac{1}{3!} \left [\left [ \left [ H_N, T^{(0)}\right ], T^{(0)}\right ],
      T^{(0)} \right ] |\Phi_0\rangle &=& 0, \;\;\;\;\;\;\;\;\; \\ 
  \langle\Phi^{pq}_{ab}|  H_{\rm N} 
   + \frac{1}{2!} \left [\left [ H_N, T^{(0)}\right], T^{(0)}\right]  
   + \frac{1}{3!} \left [\left [ \left [ H_N, T^{(0)}\right ], T^{(0)}\right ],
      T^{(0)} \right]
   + \frac{1}{4!} \left [\left [\left [\left [ H_N, T^{(0)}\right], 
     T^{(0)}\right], T^{(0)}\right], T^{(0)} \right] |\Phi_0\rangle &=& 0.
\end{eqnarray}
  \label{cceq_t0}
\end{subequations}
\end{widetext}
The states $|\Phi^p_a\rangle$ and $|\Phi^{pq}_{ab}\rangle$ are 
singly- and doubly-excited determinants obtained by replacing $one$ and $two$ 
electrons from core orbitals in $|\Phi_0\rangle$ with virtual electrons, 
respectively. And, 
$H_{\rm N} = H^{\rm DC} - \langle \Phi_0|H^{\rm DC}|\Phi_0\rangle$ is
the normal-ordered Hamiltonian.  
Similarly, the $\mathbf{T}^{(1)}$ in equation (\ref{psi-tilde}) are solutions
of the coupled nonlinear equations
\begin{widetext}
\begin{subequations}
\begin{eqnarray}
   \langle\Phi^p_a| H_{\rm N} 
     + \left [ H_N, {\mathbf T}^{(1)}\right ]
     + \left [ \left [ H_N, T^{(0)}\right ], {\mathbf T^{(1)}} \right ]
     + \frac{1}{2!} \left [ \left [ \left [ H_N, T^{(0)}\right ],
        T^{(0)}\right ], {\mathbf T^{(1)}} \right ] |\Phi_0\rangle  \nonumber \\
   =
   \langle\Phi^p_a|\left [ {\mathbf D}, T^{(0)}\right ]
   + \frac{1}{2!}\left [ \left [ {\mathbf D}, T^{(0)}\right ], T^{(0)} 
   \right ]|\Phi_0\rangle, \;\;\;\;\;\;\;\;\;\;\;\;\; \\ 
  \langle\Phi^{pq}_{ab}|  H_{\rm N} 
   +  \left [ H_N, {\mathbf T^{(1)}} \right ]
   +  \left [ \left [ H_N, T^{(0)}\right ], {\mathbf T^{(1)}} \right] 
   + \frac{1}{2!}\left [ \left [ \left [ H_N, T^{(0)}\right], T^{(0)}\right ],
      {\mathbf T^{(1)}}\right ]
   + \frac{1}{3!}  \left [ \left [ \left [ \left [ H_N, T^{(0)}\right ],
      T^{(0)}\right ],  T^{(0)}\right ], {\mathbf T^{(1)}}
   \right ]|\Phi_0\rangle \nonumber \\
   = 
   \langle\Phi^{pq}_{ab}| \left [ {\mathbf D}, T^{(0)} \right ]
   + \frac{1}{2!} \left [ \left [ {\mathbf D}, T^{(0)}\right ], T^{(0)} 
   \right ]|\Phi_0\rangle,  \;\;\;\;\;\;\;\;\;\;\;\;\;  
\end{eqnarray}
  \label{cceq_t1}
\end{subequations}
\end{widetext}
The above equations are coupled to the $T^{(0)}$ equations as these require
the values of $T^{(0)}$. We solve these equations using the Jacobi method, and
to remedy the slow convergence of this method we employ direct inversion of 
the iterated subspace (DIIS) \cite{pulay-80}. 

  In the PRCC theory, the ground state dipole polarizability of close-shell
atoms or ions is \cite{chattopadhyay-15}
\begin{equation}
  \alpha = -\frac{\langle \tilde \Psi_0|{\mathbf D}|\tilde \Psi_0\rangle}
            {\langle \tilde \Psi_0|\tilde \Psi_0\rangle}
\end{equation}
From Eq. (\ref{psi-tilde}), using the expression of  $|\tilde \Psi_0\rangle$
we can write  
\begin{equation}
  \alpha = -\frac{\langle \Phi_0|\mathbf{T}^{(1)\dagger}\bar{\mathbf{D}} + 
   \bar{\mathbf{D}}\mathbf{T}^{(1)}|\Phi_0\rangle}{\langle\Psi_0|\Psi_0\rangle},
   \label{dbar}
\end{equation}
where $\bar{\mathbf{D}}, = e^{{T}^{(0)\dagger}}\mathbf{D} e^{T^{(0)}}$, and
in the denominator $\langle \Psi_0| \Psi_0\rangle$ is the normalization factor.
Considering the computational complexity, we truncate $\bar{\mathbf{D}}$ and
the normalization to factor to second order in the cluster amplitudes. Based 
on earlier studies, the contributions from the higher order terms are 
negligible \cite{chattopadhyay-14,chattopadhyay-15}.


\section{Calculational Details}

   The use of basis set with good descriptions of single-electron
wavefunctions and energies is critical to get accurate results from RCC 
and PRCC computations. In this work use the Gaussian-type orbitals (GTOs) 
\cite{mohanty-91} as the single-electron basis for RCC and PRCC computations. 
The GTOs are finite basis set orbitals in which the orbitals are expressed as 
linear combinations of Gaussian-type functions (GTFs). Specially, the GTFs
of the large component of the orbitals have the form
\begin{equation}
  g^L_{\kappa p} (r) = C^L_{\kappa i} r^{n_\kappa} e^{-\alpha_p r^2},
\end{equation}
where $p = 0$, $1$, $2$, $\ldots$, $m$ is the GTO index and $m$ is the number 
of GTFs. And, the exponent $\alpha_p = \alpha_0 \beta^{p-1}$, where 
$\alpha_{0}$ and $\beta$ are two independent parameters optimized separately
for each orbital symmetries. This choice of the exponents is referred to as 
the even-tempered basis set. The small components of orbitals are derived
from the large components using the kinetic balance 
condition \cite{stanton-84}. To incorporate the effects of the finite size of 
the nucleus we use two-parameter finite size Fermi density distribution
\begin{equation}
   \rho_{\rm nuc}(r) = \frac{\rho_0}{1 + e^{(r-c)/a}},
\end{equation}
where, $a=t 4\ln(3)$. The parameter $c$ is the half charge radius of the 
nucleus so that $\rho_{\rm nuc}(c) = {\rho_0}/{2}$, and $t$ is the skin 
thickness. 

  To generate GTO basis we optimize $\alpha_{0}$ and $\beta$ parameters
so that the orbital energies, both the core and virtual orbitals, match the
numerical orbitals obtained from the the GRASP2K code \cite{jonsson-13}. In
addition, we also match the the self-consistent field (SCF) energies. It must
be mentioned here that the virtual orbitals $d$ (for Al$^+$ and Ga$^+$) and 
$f$ (for Ga$^+$, In$^+$, and Tl$^+$) symmetries have significant contributions
to the dipole polarizability. Hence, it is essential to optimize the 
virtual orbitals in $d$ and $f$ symmetries. The optimized $\alpha_{0}$ and 
$\beta$ parameters for all the ions are given in the Table \ref{basis}. More
detailed comparisons of the orbital energies are provided in the Appendix.

\begin{table}[h]
   \caption{The $\alpha_0$ and $\beta$ parameters of the even tempered GTO basis 
    used in our calculations.}
   \label{basis}
   \begin{ruledtabular}
   \begin{tabular}{cccccccc}
     Ion & \multicolumn{2}{c}{$s$} & \multicolumn{2}{c}{$p$} & 
     \multicolumn{2}{c}{$d$} \\
     \cline{2-3} \cline{4-5} \cline{6-7}
     & $\alpha_{0}$  & $\beta$ & $\alpha_{0}$ & $\beta$  
     & $\alpha_{0}$  & $\beta$ \\
    \hline
     ${\rm B}^+$  &\, 0.0046  &\, 2.258 &\,   &\,  &\,   &\,  \\
     ${\rm Al}^+$ &\, 0.0020  &\, 2.038 &\, 0.0020  &\, 2.105 &\,   &\,  \\
     ${\rm Ga}^+$ &\, 0.0046  &\, 2.258 &\, 0.0048  &\, 2.215 &\, 0.0045  &\,2.120 \\
     ${\rm In}^+$ &\, 0.0053  &\, 1.862 &\, 0.0052  &\, 1.870  &\, 0.0058  &\, 1.880 \\
     ${\rm Tl}^+$ &\, 0.0570  &\, 1.895 &\, 0.0498  &\, 1.820 &\, 0.0615  &\, 1.955 \\
   \end{tabular}
   \end{ruledtabular}
\end{table}

To further optimize the orbitals, we incorporate the effects of Breit 
interaction, vacuum polarization and the self-energy corrections in the 
orbital basis set. The improved orbitals are then used in RCC and PRCC 
computations. This leads to, through a change in the cluster amplitudes,
a small but important change in the dipole polarizability of all the ions. 
The detailed comparison of the orbital energy corrections from the afore
mentioned effects are given in the Appendix.


\section{Results and Discussions}

 The SCF energies computed from the optimized GTO match very well with the 
GRASP2K results for all the ions. The largest deviation is of the order 
$10^{-3}$ hartree and this is observed in the case of In$^+$. For the
remaining ions the deviation is much smaller, and lowest is in the case
of B$^+$. For which the deviation is of order $10^{-6}$ hartree. A detailed
comparison of the SCF and orbital energies is provided in the Appendix. In the
Table \ref{alpha_conv} we show the convergence pattern of $\alpha$ with the 
basis size. As computations with the DCB Hamiltonian are more compute 
intensive, to determine the converged basis we use the DC Hamiltonian. For 
example, the computation of $\alpha$ for heavy ions like Tl$^+$ with moderate 
basis set size of 154 orbitals needs a few weeks to complete. As discernible from the 
table, we start with a moderate basis size orbitals and add orbitals 
systematically in each symmetry until $\alpha$ converges up to $10^{-3}$ 
atomic units or smaller. For a better description the results of $\alpha$ and 
observed trends of the correlation effects are discussed for each of the ions 
separately.
\begin{table}
  \caption{Convergence pattern of $\alpha$ calculated using Dirac-Coulomb 
           Hamiltonian as function of the basis set size. The values listed are 
           in atomic units ($a_0^3$).}
  \label{alpha_conv}
  \begin{ruledtabular}
  \begin{tabular}{lcr}
      No. of orbitals & Basis size & $\alpha $   \\
      \hline
              \multicolumn{3}{c}{B$^+$}\\
      107 & $(13s, 13p, 12d, 7f, 6g, 4h)    $& 9.292  \\
      129 & $(15s, 15p, 14d, 9f, 8g, 6h)    $& 9.346  \\
      151 & $(17s, 17p, 16d, 11f, 10g, 8h)  $& 9.358  \\
      168 & $(20s, 20p, 19d, 13f, 12g, 10h)  $ & 9.413  \\
      173 & $(21s, 21p, 20d, 13f, 12g, 10h)  $ & 9.413  \\
      
              \multicolumn{3}{c}{Al$^+$}\\
      131 & $(19s, 19p, 11d, 9f,  9g,  8h) $ & 23.618   \\
      148 & $(22s, 22p, 12d, 10f, 10g, 9h) $ & 23.652   \\
      159 & $(23s, 23p, 13d, 11f, 11g, 10h) $ & 23.789  \\
      166 & $(24s, 24p, 14d, 12f, 11g, 10h) $ & 23.999  \\
      169 & $(25s, 25p, 14d, 12f, 11g, 10h) $ & 23.999  \\
              \multicolumn{3}{c}{Ga$^+$}\\
      116 & $(16s, 16p, 14d, 7f,  7g,  6h) $ & 18.050  \\
      132 & $(18s, 18p, 16d, 8f,  8g,  7h) $ & 18.050  \\
      152 & $(20s, 20p, 18d, 10f, 10g, 8h) $ & 18.053  \\
      172 & $(22s, 22p, 20d, 12f, 11g, 10h) $ & 18.056  \\
      177 & $(23s, 23p, 21d, 12f, 11g, 10h) $ & 18.056  \\
              \multicolumn{3}{c}{In$^+$}\\
      123 & $(20s, 20p, 15d, 8f,  6g,  5h) $ & 24.748  \\
      139 & $(21s, 21p, 16d, 9f,  7g,  6h) $ & 24.746  \\
      150 & $(22s, 22p, 17d, 10f, 8g,  7h) $ & 24.744  \\
      162 & $(22s, 22p, 17d, 12f, 10g, 10h) $ & 24.744  \\
              \multicolumn{3}{c}{Tl$^+$}\\
      134 & $(16s, 15p, 15d, 12f, 9g, 8h) $ &   20.026  \\
      147 & $(17s, 16p, 16d, 13f, 10g, 10h) $ & 20.173 \\
      156 & $(18s, 17p, 17d, 14f, 11g, 10h) $ & 20.129  \\
      161 & $(19s, 18p, 18d, 14f, 11g, 10h) $ & 20.215 \\
      171 & $(21s, 20p, 20d, 14f, 11g, 10h) $ & 20.216  \\
  \end{tabular}
  \end{ruledtabular}
\end{table}


\subsection{B$^+$ and Al$^+$}

\begin{figure*}
 \includegraphics[height=17cm, angle=-90]{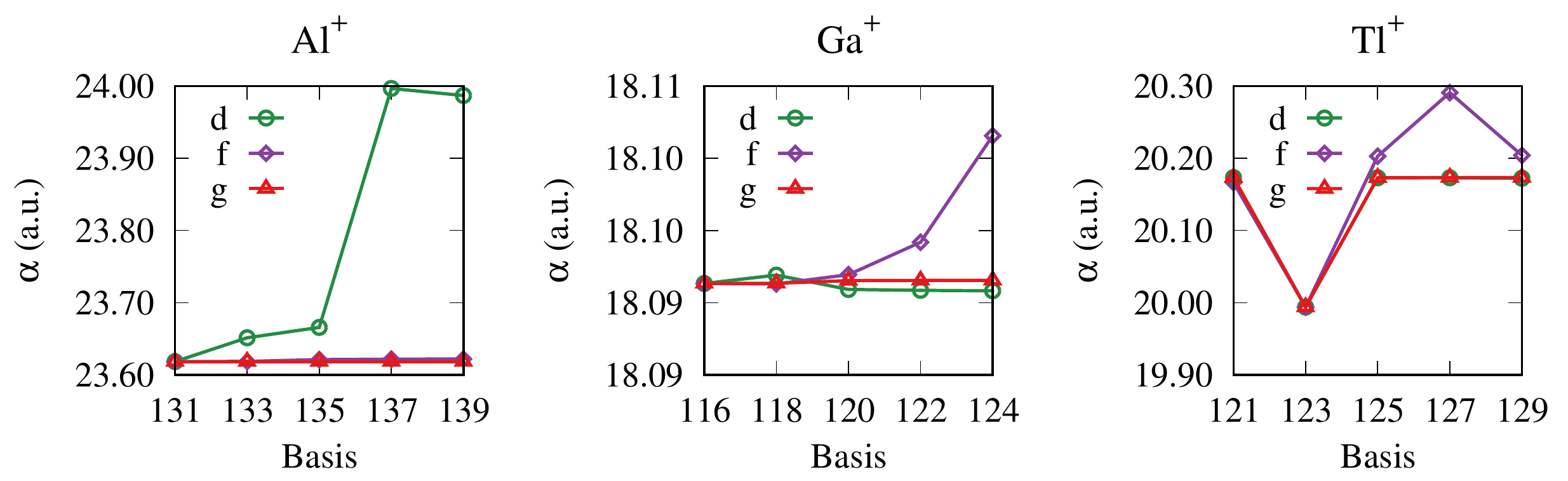}
 \caption{The trend of contributions to $\alpha$ from $d$, $f$ and $g$ virtual orbitals.}
 \label{basis_vs_alpha}
\end{figure*}

From the results given in Table \ref{alpha_conv}, we find that the change is below
$10^{-3}$ in $\alpha$ for B$^+$ when the basis set is increased from 168 
to 173. So, to minimize the computation time, we consider the basis set with 
168 orbitals as optimal, and use it for the further analysis. Similarly, for
Al$^+$ we find that a basis set with 166 orbitals is optimal. To analyse the 
symmetry wise contributions from the virtual orbitals, we plot the values of 
$\alpha$ for each symmetry with respect to the basis size in 
Fig. \ref{basis_vs_alpha}. From the figure it evident that the $d$ virtual 
orbitals have significant contributions for Al$^+$. We attribute this to the 
correlation effects arising from the strong mixing of the $2p$ core electrons 
with $d$ virtual electrons. This indicates that to get high quality results
for Al$^+$ it is essential to optimize the $d$ virtual orbitals.

 The value of $\alpha$ computed from the converged basis set is listed in 
Table \ref{alpha_final}, for comparison the results reported in previous 
works are also listed. For B$^+$, as we see from the table, there are very
few theoretical results in literature. And, to best of our knowledge, there
are no experimental data. From the previous works the average value reported 
is 9.6. However, our LPRCC result of 12.809, is $\approx 33$\% larger than the 
previous theory results and, therefore, indicates missing of important 
correlation effects. This is in contrast to the trends observed in our previous 
works \cite{chattopadhyay-12a, chattopadhyay-12b, chattopadhyay-13a, 
chattopadhyay-13b,chattopadhyay-14,chattopadhyay-15}, where LPRCC 
provides a reliable result for the ground state polarizability. This could be
due to the stronger electron-correlation effects associated with the 
two-valence nature of the ions. And, which is further enhanced due to the 
orbital contraction as these are singly charged ions. From our calculations
we find that the nonlinear terms in PRCC theory are very important and must be 
included in the computation of $\alpha$ for such a complicated ions.
In particular, we find that a  nonlinear diagram 
arising from the PRCC term $ \contraction[0.3ex] {}{H}{_N}{T} 
\contraction[0.6ex] {}{H}{_N T^{(0)}}{T} H_N T^{(0)}_2 {\mathbf T^{(1)}_1}$,
has a very large contribution but with opposite sign. We  observe this trend
for all the ions in the group. Our PRCC and PRCC(T) results, 9.413 and 9.415, 
respectively, are in good agreement. However, these are slightly lower than the 
previous results. This difference can be attributed to the effects arising 
from the inclusion Breit interaction in our computations.  

\begin{figure*}
 \includegraphics[height=17cm, angle=-90]{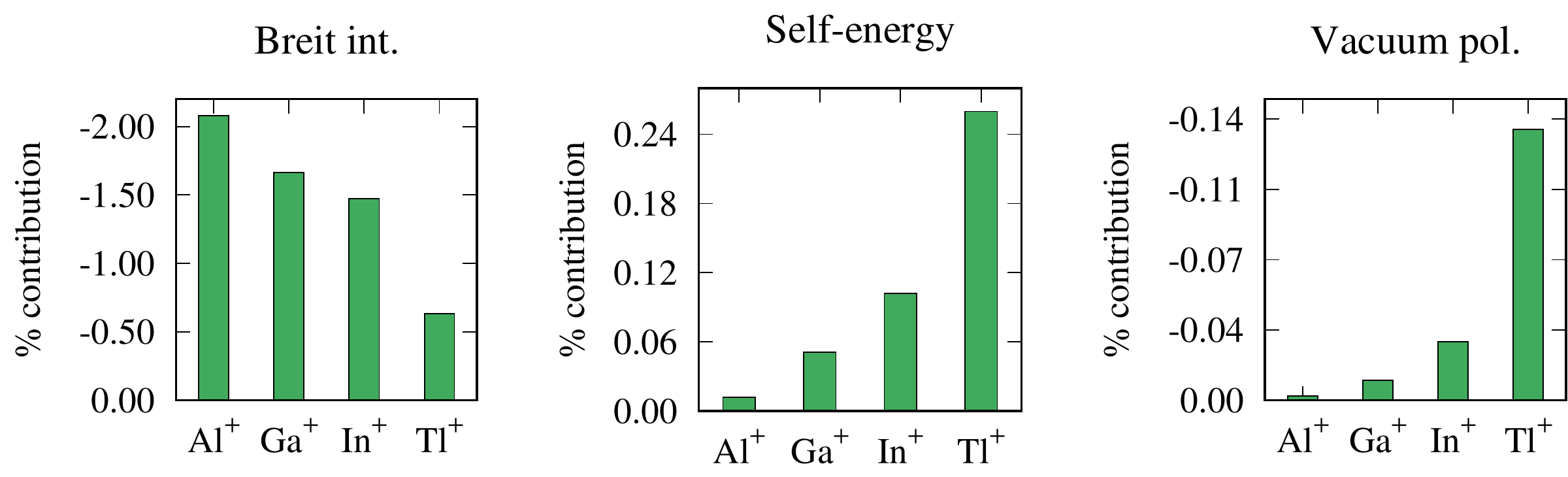}
 \caption{The percentage contributions from the Breit interaction, the 
         vacuum polarization and the self-energy corrections.}
 \label{BSVcontribution}
\end{figure*}

For Al$^+$, among all the ions considered it has the largest number of 
previous results. However, like in the case of B$^+$ all of these are 
theoretical results. Except for the RCI based finite-field results
\cite{yu-13}, the values of $\alpha$ reported in previous works
are very close to each other. The average value from the reported values is 
24.1. The value 23.78 reported in Ref. \cite{yu-13} is on the lower 
side than the others. In terms of the many-body theory method the method used
in K\'allay {\it et al.} \cite{kallay-11} is closest to ours. They have used 
the relativistic general-order coupled-cluster theory, where the higher-order 
cluster excitations are considered by using the many-body diagrammatic 
techniques based automated programming tools.  However, one important 
difference is that they use DC Hamiltonian, where as we use the DCB Hamiltonian.
In fact, this is the main reason for our PRCC and PRCC(T) results of 23.502 and 
23.516, respectively,  to be lower than that of Ref.\cite{kallay-11}, and 
others. This is evident from the Table \ref{alpha_sepa} where we list 
contributions from different effects. Our result of 23.999 based on the
DC Hamiltonian is in excellent agreement with the previous values. As we 
observe from the table, the contribution from the Breit interaction is -0.499, 
which reduces the DC result of $\alpha$. The contributions from the vacuum 
polarization and the self-energy correction are -0.0004 and 0.0028, 
respectively. As shown in the Fig. \ref{BSVcontribution}, in terms of percentage 
these are $\approx -0.002$\% and $\approx 0.01$\%, respectively.  
An important point to observe here is that, the contribution from the vacuum 
polarization is opposite in phase to that of the self-energy correction. 
Interestingly, we observed the same pattern 
in our previous work on dipole polarizability of group-IIB 
elements \cite{chattopadhyay-15}.  The inclusion of triples perturbatively 
improves the result further, and the contribution is 0.014. 

  In the Table \ref{alpha_termwise} we provide the term wise contributions to 
$\alpha$.  As we observe, for both the ions, the leading order (LO) term is
${\bf T_1^{(1)\dagger}D +  H.c.}$. Quantitatively, it is $\approx 106$\% 
and 105\% of the total value in the case of B$^+$ and Al$^+$, respectively.
This is expected as it subsumes the contributions from the Dirac-Hartree-Fock 
and the core polarization effects. The next to leading order (NLO) term is 
${\bf T_1{^{(1)\dagger}} D T_2^{(0)} + H.c.}$, which has a contribution of 
$\approx -3$\% of the total value in both the ions. The last two dominant 
contributions are from ${\bf T_2{^{(1)\dagger}} DT_2^{(0)}}$
and ${\bf T_1{^{(1)\dagger}} DT_1^{(0)}}$ terms, respectively.
The contribution from the former is roughly twice than the latter but 
opposite in phase.

\begin{table}
    \caption{Separate contributions to $\alpha$ from different interaction terms in 
            the Hamiltonian used in PRCC calculations.} 
    \label{alpha_sepa}
    \begin{center}
    \begin{ruledtabular}
    \begin{tabular}{lcccc}
        Method &  \multicolumn{1}{r}{$\rm{Al^+}$}
        & \multicolumn{1}{r}{$\rm{Ga^+}$}  
        & \multicolumn{1}{r}{$\rm{In^+}$}  
        & \multicolumn{1}{r}{$\rm{Tl^+}$}  \\
        \hline
        PRCC(DC) 
        &       23.9989     &  18.0556   &  24.7449    &   20.2159 \\ 
        Breit int.
        &      -0.4994      &  -0.3006   &  -0.3647    &  -0.1283 \\
        Vacuum pol.
        &    -0.0004   &  -0.0018   &  -0.0070    &       -0.0274 \\
        Self-energy
        &     0.0028    &   0.0090   &   0.0249    &    0.0526 \\
        Total    
        &  23.5019      &  17.7621   &  24.3981   &     20.1128 \\
    \end{tabular}
    \end{ruledtabular}
    \end{center}
\end{table}


\subsection{Ga$^+$ and In$^+$}

  From our results we find that for Ga$^+$ both the $d$ and $f$ virtual 
orbitals have significant contribution to $\alpha$. This is due to the 
mixing of these virtual orbitals of $d$ and $f$ symmetries with the $3p$ and 
$3d$ core orbitals, respectively. Similarly, for In$^+$ there is a mixing 
between the $4d$ core electrons and $f$ virtual orbitals. In 
Table \ref{alpha_final} we provide the converged value of $\alpha$ from our 
computations and compare with the data available in the literature. As we 
can observe from the table, there are only two and four previous results
in the case of Ga$^+$ and In$^+$, respectively.  All of these are theoretical
results and to the best of our knowledge, there are no experimental results
of $\alpha$ for these two ions. For Ga$^+$, the value of $\alpha$, 18.14, 
from Ref. \cite{reshetnikov-08} using sum-rule is higher than the 
CICP value of 17.92 \cite{cheng-13}. Our PRCC (DC) result of 18.056, listed
in the Table \ref{alpha_sepa}, lies between these two results. Our PRCC result
of 17.762 using DCB Hamiltonian is lower than both the previous results. 
As discussed before, this difference can be attributed to Breit interaction, 
which has the contribution of -0.3006. The PRCC (T) result of 17.814 shows a 
marginal increase and the contribution from the perturbative triples is 0.052. 
The contributions from the vacuum polarization and  the self-energy correction 
are -0.0018 and 0.0090, respectively. Like in Al$^+$, these are of opposite 
phases, and are larger in magnitude by $\approx 350$\% and $\approx 221$\%, 
respectively than Al$^+$. Interestingly, the Breit contribution 
to $\alpha$ is smaller,  by $\approx 66$\%, than the Al$^+$. 

 For In$^+$, there is a large variation in the value of $\alpha$  reported in
the previous works. The lowest value of 18.8 is obtained by using the sum-rule
\cite{reshetnikov-08}, while the highest value 24.33 is based on the
finite-field method \cite{yu-15}. The value of 24.01 reported based on 
CI + all-order method \cite{safronova-11b} lies between the two previous 
results. Our PRCC and PRCC(T) results are 24.398 and 24.467, respectively. 
These values are higher than the previous results. Quantitatively, our PRCC
result is larger by $\approx 1.6$\% and $\approx 0.3$\% than the values in 
Refs. \cite{safronova-11b} and \cite{yu-15}, respectively. The contributions 
from the Breit interaction, vacuum polarization, self-energy correction and 
the perturbative triples are -0.3647, -0.0070, 0.0249 and 0.069, respectively. 
In terms of percentage, these are $\approx 1.5$\%, $\approx 0.03$\%, $\approx 0.1$\%
and $\approx 0.3$\%, respectively of the DC value of $\alpha$.
The contributions are larger in magnitude by $\approx 21 $\%, $\approx 288$\%, 
$\approx 176$\% and $\approx 32$\% than the Ga$^+$ ion. This clearly indicates
the importance of incorporating these effects to obtain accurate results.

 Examining the term wise contributions, both ions follow the trend of B$^+$ and 
Al$^+$, where the LO and NLO terms are ${\bf T_1^{(1)\dagger}D +  H.c.}$ and
${\bf T_1{^{(1)\dagger}} D T_2^{(0)} + H.c.}$, respectively. While the
LO contributes $\approx 109$\%, NLO contributes $\approx -4$\% of the total 
value each for both the ions. The next two dominant contributions are from
${\bf T_2{^{(1)\dagger}} DT_2^{(0)}}$ and ${\bf T_1{^{(1)\dagger}} DT_1^{(0)}}$ 
terms, respectively. The contribution from 
${\bf T_2{^{(1)\dagger}} DT_2^{(0)}}$ is $\approx 2.8$\% each for Ga$^+$ and 
In$^+$ ions. And, the contribution from ${\bf T_1{^{(1)\dagger}} DT_1^{(0)}}$ 
is $\approx -1.7$\% each in the case Ga$^+$ and In$^+$.

\begin{table}
  \caption{The final value of $\alpha$ (in a.u.) from our calculations are 
          compared with the other theory and experimental results. The values
          of $\alpha$ listed from present work include the effects of Breit 
          interaction, vacuum polarization and the self-energy corrections.} 
  \label{alpha_final}
  \begin{center}
  \begin{ruledtabular}
  \begin{tabular}{lllll}
    \multicolumn{1}{c}{Ion}      & \multicolumn{1}{c}{Present}  &
    \multicolumn{1}{c}{Method}   & \multicolumn{1}{c}{Previous} &
    \multicolumn{1}{c}{Method}                              \\
                                 & \multicolumn{1}{c}{work}     &
                                 & \multicolumn{1}{c}{works}    & \\
    \hline
    $\rm{B^+}$& 12.809 & LPRCC & 9.62 \text{\cite{safronova-11b}} & CI+all-order \\
              & 9.413  & PRCC &  9.57 \text{\cite{archibong-90}} & CCD+ST(CCD)    \\
              & 9.415  & PRCC(T) & 9.64(3) \text{\cite{cheng-12}} & CICP    \\ \\ 
    \hline

   $\rm{Al^+}$& 28.624 & LPRCC & 24.05 \text{\cite{safronova-11b}} & CI+all-order \\
              & 23.502 & PRCC  & 24.14(8) \text{\cite{kallay-11}} & RCC   \\
              & 23.516 & PRCC(T) & 23.78(15)\footnotemark[1] \text{\cite{yu-13}} & Finite-field  \\
              &        &         & 24.07(41)\footnotemark[2] \text{\cite{yu-13}} & Finite-field  \\
              &        &         & 24.12 \text{\cite{hamonou-08}} & CI   \\
              &        &         & 24.14(12) \text{\cite{mitroy-09}} & CICP   \\
              &        &         & 24.20(75) \text{\cite{reshetnikov-08}} & Sum-rule   \\ 
              &        &         & 24.2 \text{\cite{archibong-91}} & MBPT   \\ \\ 

    \hline

   $\rm{Ga^+}$& 21.722 & LPRCC & 17.95 \text{\cite{cheng-13}} & CICP     \\
              & 17.762 & PRCC & 18.14(44) \text{\cite{reshetnikov-08}} & Sum-rule \\
              & 17.814 & PRCC(T) &   &           \\ \\
                         
    \hline

   $\rm{In^+}$ & 30.167 & LPRCC & 24.01 \text{\cite{safronova-11b}} & CI+all-order \\
              & 24.398  & PRCC & 24.16(3)\footnotemark[3] \text{\cite{yu-15}} &Finite-field \\
              & 24.467  & PRCC(T) & 24.33(15)\footnotemark[4] \text{\cite{yu-15}}&Finite-field \\              
              &         &    & 18.8(13) \text{\cite{reshetnikov-08}} & Sum-rule  \\ \\               
              
    \hline

    $\rm{Tl^+}$&22.834  & LPRCC & 19.60 \text{\cite{zuhrianda-12}} & CI+all-order     \\
              & 20.113  & PRCC & 12.7(12) \text{\cite{reshetnikov-08}} & Sum-rule  \\
              & 20.129 & PRCC(T) &  &  \\
  \end{tabular}
  \end{ruledtabular}
  \end{center}
  \footnotetext[1]{ Finite-field using energies from RCI calculations.}
  \footnotetext[2]{ Finite-field using energies from RCC calculations.}
  \footnotetext[3]{ Finite-field using energies from RCI calculations.}
  \footnotetext[4]{ Finite-field using energies from RCC calculations.}
\end{table}

\begin{table}
    \caption{Contribution to $\alpha$ (in a.u.) from different terms and their
             hermitian conjugates in PRCC theory.}
    \label{alpha_termwise}
    \begin{center}
    \begin{ruledtabular}
    \begin{tabular}{lccccc}
        Terms + h.c. & \multicolumn{1}{r}{$\rm{B^+}$}
        & \multicolumn{1}{r}{$\rm{Al^+}$}
        & \multicolumn{1}{r}{$\rm{Ga^+}$}
        & \multicolumn{1}{r}{$\rm{In^+}$}
        & \multicolumn{1}{r}{$\rm{Tl^+}$}  \\
        \hline
        $\mathbf{T}_1^{(1)\dagger}\mathbf{D} $
        & 9.9782   &  25.2392   &    19.7942 &  27.0396   &    22.3619    \\
        $\mathbf{T}_1{^{(1)\dagger}}\mathbf{D}T_2^{(0)} $
        & -0.3162   & -0.7448    & -0.6858  &  -1.0204   &     -1.1296    \\
        $\mathbf{T}_2{^{(1)\dagger}}\mathbf{D}T_2^{(0)} $
        &  0.2740   &  0.6074    &  0.5088  &  0.6818   &       0.5700    \\
        $\mathbf{T}_1{^{(1)\dagger}}\mathbf{D}T_1^{(0)} $
        & -0.1804   &  -0.3342    & -0.3222  &  -0.4134   &    -0.1352    \\
        $\mathbf{T}_2{^{(1)\dagger}}\mathbf{D}T_1^{(0)} $
        &  0.0106   &  0.0206    &   0.0170  &  0.0212   &     -0.0078    \\
        Normalization & 1.0367  & 1.0329   & 1.0696  &  1.0632   &    1.0714 \\
        Total         & 9.4207 & 23.9989  & 18.0556 &  24.7449   &    20.2159 \\
    \end{tabular}
    \end{ruledtabular}
    \end{center}
\end{table}


\subsection{Tl$^+$}

  The orbital energies obtained from the GTO are in very good agreement with 
those of numerical orbitals from GRASP2K. The corrections from the 
self-energy are consistent with the trend observed in the other 
ions. A detailed comparison and discussions of the orbital energies and
various corrections are given in the Appendix. The corrections from the Breit 
interaction and vacuum polarization requires special 
attention. The correction from the Breit interaction, 
$\Delta\epsilon_{\rm Br}$, has a different trend than the previous ions. And,
that is the negative sign of $\Delta\epsilon_{\rm Br}$ for 
$4f_{5/2}$, $4f_{7/2}$ and $5d_{5/2}$ orbitals. Interestingly, the same 
pattern was also observed in the case of Hg in our previous work 
\cite{chattopadhyay-15} and, as mentioned there, it may be due to the large 
weight factor $2j + 1$ associated with the two-electron exchange integrals.

  Like in  In$^+$ and other ions, $\Delta\epsilon_{\rm Ue}$ is negative for 
all the $s_{1/2}$ orbitals. There is, however, a difference in the phase for 
$p_{1/2}$ orbitals where, unlike the previous ions, $\Delta\epsilon_{\rm Ue}$ 
is negative for all the $p_{1/2}$ orbitals. This trend of negative 
$\Delta\epsilon_{\rm Ue}$ is consistent with the case of Hg and Ra$^{2+}$ 
observed in our previous works Refs. \cite{chattopadhyay-15} and 
\cite{chattopadhyay-13b}, respectively. We attribute this to the larger 
relativistic effects due to the stronger nuclear potential in these ions.

  The converged value of $\alpha$ is computed by using the optimal basis 
of 161 orbitals. The computation, like in the previous ions, is incorporating 
the effects of the Breit interaction, vacuum polarization, and the self-energy 
corrections. As shown in the Fig. \ref{basis_vs_alpha} the virtual orbitals of 
the $f$ symmetry have dominant contributions and are important to get the 
converged value of $\alpha$. This arises from the large mixing between the 
$f$ virtual orbitals and $5d$ core orbitals.

In the Table \ref{alpha_final} we compare our converged result with
the previous values. For comparison we found only two theoretical results
from the literature. However, there is a large variation in these two results,
the value of 19.60 obtained from CI + all-order theory \cite{zuhrianda-12} is $\approx 54$\% 
larger than the value of 12.7 obtained from the sum-rule method \cite{reshetnikov-08}. 
Our PRCC and PRCC(T) results of 20.113 and 20.129, respectively are in 
good agreement with the Ref. \cite{zuhrianda-12}. However, our LPRCC result 
of 22.834 is $\approx 17$\% larger than the Ref. \cite{zuhrianda-12}. The reason for 
this is similar to the case of the previous ions--large cancellation due to the 
contribution from a non-linear diagram arising from the PRCC 
term $ \contraction[0.3ex] {}{H}{_N}{T}
\contraction[0.6ex] {}{H}{_N T^{(0)}}{T} H_N T^{(0)}_2 {\mathbf T^{(1)}_1}$. 
The contribution from the Breit interaction is $-$0.1209. This has the 
same phase as in the previous ions but smaller by $\approx 66$\% than
In$^+$. The contributions from the vacuum polarization and the self-energy 
correction are $-$0.0258 and 0.0495, respectively. These are also in the same 
phase as the previous ions but larger by  $\approx 69$\% and  98\%, 
respectively than In$^+$. We also observe that the term wise contributions
are also of the same pattern as the previous ions.


To estimate the upper bound on the uncertainty associated with the value 
of $\alpha$ in the present work we have isolated four different sources. 
While some of these have very small contributions and can be neglected, some
are must to be included. 
The first source of uncertainty is due to the truncation of the basis set in our 
computations. Since the values of $\alpha$ reported in Table \ref{alpha_final}
are using the converged basis set where the change in $\alpha$ is of the order 
of $10^{-3}$ or less, we neglect this uncertainty also.
The second source of uncertainty is due to the truncation of the dressed 
operator $\bar{\mathbf{D}}$ in equation (\ref{dbar}) to include the
cluster operators up to the second order only. In our previous work \cite{mani-10}, 
using an iterative scheme, we have shown that the contribution from the third- and 
higher-order terms is negligible and therefore we can neglect this uncertainty.
The third source of uncertainty is due to the partial inclusion of 
triples (${\mathbf T}^{(1)}_3$) in the PRCC theory. The maximum contribution from 
the perturbative triples is $\approx 0.3$\% in the case of Ga$^+$. Since the
perturbative scheme accounts for the dominant contributions, we can assume
$0.3$\% as an upper bound to this uncertainty.
The last source of uncertainty in our computation is associated with the
frequency-dependent Breit interaction which we do not include in the present work.
However, in our previous work \cite{chattopadhyay-14} using a series of
calculations using GRASP2K we estimated an upper bound to be $0.13$\% in the case 
of Ra. Since Ra has higher $Z$ than Tl, in this work
as well we assume $0.13$\% as an upper bound on the uncertainty due to 
frequency-dependent Breit interaction. 
Combining all these sources of uncertainties, the upper bound on the uncertainty 
in the value of $\alpha$ is 0.5\%.


\section{Conclusion}

  We have computed the ground state electric dipole polarizability 
group-IIIA ions using the PRCC theory. To account for the relativistic effects 
and QED corrections, we have used the Dirac-Coulomb-Breit Hamiltonian with 
the corrections from the Uehling potential and the self-energy. The effects 
of triple excitations are considered perturbatively.
Our results from PRCC and PRCC(T) using the Dirac-Coulomb Hamiltonian 
are in excellent agreement with the previous results for all the ions. The
results using the Dirac-Coulomb-Breit Hamiltonian are, however, lower than 
most of the previous results except for In$^+$. We attribute this to the 
effects of the Breit interaction, which is considered in our work but not 
in the previous works. The other important observation from our computations
is that, we need to go beyond the LPRCC to obtain accurate results for 
the group-IIIA ions. The LPRCC results are on an average $\approx 24$\% 
larger than the PRCC results. This could be due to the strong correlation 
effects arising from the divalent nature of the group-IIIA ions. And, to 
account for such a large correlation effects the nonlinear terms in PRCC 
theory must be included.

Based on our analysis of the corrections arising from the Breit interaction 
we find two trends. First, the contribution for all the ions are negative, 
and hence reduces the total value of $\alpha$. The same pattern of is also 
observed in the case of noble-gas atoms \cite{chattopadhyay-12b}, 
alkaline-earth-metal atoms \cite{chattopadhyay-14} and the group-IIB 
elements \cite{chattopadhyay-15}. In the case of singly ionized 
alkali-metal atoms \cite{chattopadhyay-13a}, however, we get a different
trend. Second, in terms of the percentage contribution, we observed the 
largest contribution of $\approx 2.1$\% in the case of Al$^+$. 
And, as we go towards the heavier ions the contribution decreases, the lowest 
is $\approx - 0.63$\%, in the case of Tl$^+$. A similar pattern is also observed in 
the case of alkaline-earth-metal atoms and the group-IIB elements where heavier 
atoms Ra and Hg have the smaller contributions, of 
$\approx -0.4$\% and $-0.02$\%, respectively, than the lighter ones.  In the 
case of noble-gas atoms, however, we observed an opposite trend where the 
heaviest atom Rn has the largest contribution of $\approx 0.1$\%. 

For the Uehling potential and the self-energy corrections, we observed a 
trend of increasing contributions from the lighter ions to the heavier ions. 
This is to be expected as the heavier atoms have the larger relativistic 
effects. The largest contributions are $\approx -0.1$\% and $0.3$\% from the 
Uehling potential and the self-energy corrections, respectively, in the case 
of Tl$^+$.  We observed an opposite trend from the Uehling potential in the 
case of the group-IIB elements \cite{chattopadhyay-15}, where Zn has larger 
contribution ($\approx -0.3$\%) than the Hg ($\approx -0.1$\%).  For the 
self-energy correction, the group-IIB elements show a mix behavior where both 
Zn and Hg have larger contribution than Cd.


\begin{acknowledgments}
We thank Chandan Kumar Vishwakarma for useful discussions. The results 
presented in the paper are based on the computations using the HPC 
facilities at the Indian Institute of Technology Delhi, New Delhi and
Physical Research Laboratory, Ahmedabad.
\end{acknowledgments}

\appendix


\section{Breit interaction, vacuum polarization and self energy corrections}

   For Breit interaction we use the approach introduced by Grant and 
Pyper \cite{grant-76} where the Breit interaction operator $g^B_{r_{12}}$ is 
expanded as a linear combination of irreducible tensor operators. We employ 
the  expressions given in Ref. \cite{grant-80} to incorporate the effects of 
$g^B_{r_{12}}$ in single-electron basis as well as RCC and PRCC calculations.  
To analyze the effects of $g^B_{r_{12}}$ in detail, we compute contributions to 
SCF energy as well as the single-electron energies for all the ions. The 
correction to single-electron energy due to Breit interaction is 
\begin{equation}
  \Delta\epsilon_{\rm Br (i)} = \epsilon^\prime_i - \epsilon_i, 
\end{equation}
where $\epsilon_i$ and $\epsilon^\prime_i$ represent the orbital
energies obtained by solving the Dirac-Hartree-Fock and 
Dirac-Hartree-Fock-Breit orbital equations self-consistently, respectively. 
Similarly, the correction to the SCF energy is
\begin{equation}
  \Delta E^{\rm SCF}_{\rm Br} = E^{\rm SCF}_{\rm DCB} - E^{\rm SCF}_{\rm DC}, 
\end{equation}
where, $E^{\rm SCF}_{\rm DCB}$ and $E^{\rm SCF}_{\rm DC}$ are the SCF 
energies computed using DCB and DC Hamiltonian, respectively.
The $\Delta E^{\rm SCF}_{\rm Br}$ computed from our implementation on GTO is 
given in Table \ref{scf_e} where we compare our results with a recently 
reported code for B-spline basis by Zatsarinny {\it el al.} \cite{oleg-16}. 
The contributions to the orbital energies are tabulated in the 
Tables \ref{orbe_btoin} (for B$^+$, Al$^+$, Ga$^+$, and In$^+$) and 
\ref{orbe_tl} (for Tl$^+$) for a quantitative description. 

\begin{table*}[t]
\caption{The SCF energy from GTO using the Dirac-Coulomb 
         Hamiltonian is compared with the GRASP2K \cite{jonsson-13} results. 
         The contributions from the Breit interaction and the vacuum
         polarization are compared with the results from 
         the B-spline code \cite{oleg-16}. All the values are in Hartree.}
\label{scf_e}
\begin{center}
\begin{ruledtabular}
\begin{tabular}{ldddddr}
    Ion & \multicolumn{2}{c}{$E^{\rm SCF}$}
        & \multicolumn{2}{c}{$\Delta E^{\rm SCF}_{\rm Br}$} 
        & \multicolumn{2}{c}{$\Delta E^{\rm SCF}_{\rm Ue}$} \\ 
    \cline{2-3}
    \cline{4-5}
    \cline{6-7}
        & {\rm GTO}  & {\rm GRASP2K} & 
          {\rm GTO} & {\rm B-spline} &
          {\rm GTO} & {\rm B-spline}  \\
    \hline 
    ${\rm B}^+$  &   -24.24516  &    -24.24516 &0.00148 & 0.00148 & -0.00004 & 0.00005 \\
    ${\rm Al}^+$ &  -242.12904  &   -242.12905 &0.04221 & 0.04222  & -0.00170 & -0.00178\\
    ${\rm Ga}^+$ & -1942.36249  &  -1942.36368 &0.85161 & 0.85207 & -0.05758 & -0.05990\\
    ${\rm In}^+$ & -5880.24254  &  -5880.24386 &4.12196 & 4.12552 & -0.39075 & -0.40518\\
    ${\rm Tl}^+$ & -20274.62436 & -20274.62463 &23.65940& 23.69367 & -4.11174& -4.23753\\
\end{tabular}
\end{ruledtabular}
\end{center}
\end{table*}

To incorporate the effects of vacuum polarization we consider the expression
given by Johnson {\em et al.} \cite{johnson-01}, where the Uehling 
potential \cite{uehling-35} is generalized to incorporate the effects of 
the finite size of the nucleus. In our previous work \cite{chattopadhyay-13b} 
we had discussed the details of the implementation. To quantify the effects 
in the present work, we compute the corrections to orbital energies as well 
as the SCF energy for all the ions. The correction to orbital energy is 
\begin{equation}
\Delta\epsilon_{\rm Ue (i)} = \epsilon^\prime_i - \epsilon_i,
\end{equation}
where $\epsilon^\prime_i$ and $\epsilon_i$ are the energies with and without
Uehling potential, respectively. Similarly, the correction to SCF energy is
\begin{equation}
  \Delta E^{\rm SCF}_{\rm Ue} = E^{\rm SCF}_{\rm (DC + Ue)} 
                              - E^{\rm SCF}_{\rm DC}, 
\end{equation}
where, $E^{\rm SCF}_{\rm DC+Ue}$ and $E^{\rm SCF}_{\rm DC}$ are the SCF 
energies computed using DC plus Uehling potential and DC Hamiltonian, 
respectively. The $\Delta E^{\rm SCF}_{\rm Ue}$ from our computations are 
tabulated and compared with the results from the B-spline code \cite{oleg-16} 
in the Table \ref{scf_e}. And, $\Delta\epsilon_{\rm Ue (i)}$ are given in 
the Tables \ref{orbe_btoin} and \ref{orbe_tl}.

The effects of the self-energy (SE) correction to orbitals are considered 
through the model Lamb-shift operator introduced by 
Shabaev {\em et al.} \cite{shabaev-13}. For this we use the code 
QEDMOD \cite{shabaev-15}, developed by the same authors, to compute
the corrections to the orbital energies. These corrections to energies 
are applied and used in the RCC and PRCC computations. A similar analysis 
was reported for the group-IIB elements in our previous 
work \cite{chattopadhyay-15}. The data on ${\rm SE}$ corrections to orbital 
energies, computed using QEDMOD code, are listed in the 
Tables \ref{orbe_btoin} and \ref{orbe_tl}.


\section{SCF and orbital energies}

 In Table \ref{scf_e} we compare the SCF energy from GTO with GRASP2K 
\cite{jonsson-13}. Similarly, the  orbital energies of GTO are tabulated and 
compared with energies of the numerical orbitals obtained from GRASP2K in the 
Tables \ref{orbe_btoin} and \ref{orbe_tl}. Considering the Breit correction, 
the sign of $\Delta E^{\rm SCF}_{\rm Br}$ is positive for all the ions and 
matches with the B-spline results. The positive sign of 
$\Delta E^{\rm SCF}_{\rm Br}$ indicates an increase in the SCF energy, which 
we attribute to the spatial contraction of the orbitals. Interestingly, we
reported the observation of the same trend of $\Delta E^{\rm SCF}_{\rm Br}$
in the case of the noble-gas \cite{chattopadhyay-12b} and group-IIB 
elements \cite{chattopadhyay-15}. Examining the values listed in the table,
we find that our GTO results are in excellent  agreement with the B-spline 
results. The largest difference is of the order of $10^{-2}$ hartree, which
occurs in the case Tl$^+$. The last two columns of Table \ref{scf_e} show 
the comparison of $\Delta E^{\rm SCF}_{\rm Ue}$ from GTO with the B-spline 
data. Unlike $\Delta E^{\rm SCF}_{\rm Br}$, $\Delta E^{\rm SCF}_{\rm Ue}$ 
from GTO has negative value for all the ions, indicating a decrease in the 
SCF energy. This decrease in SCF energy implies the relaxation of the 
orbitals. There is a sign mismatch in the results of B$^+$, though,
the contribution is very small. For the remaining ions, both sign as well as 
magnitude of GTO results are in good agreement with the B-spline data. For 
better comparison, the results from each ions are discussed separately.
\begin{table*}
        \caption{The orbital energies (in Hartree) from GTO compared with the 
                GRASP2K\cite{jonsson-13} results for B$^+$, Al$^+$, Ga$^+$, 
                and In$^+$. The contributions from the Breit interaction, vacuum
                polarization and the self-energy corrections to GTOs
                are also listed. The self-energy corrections are calculated 
                using the code QEDMOD by Shabaev {et al.} \cite{shabaev-15}.  
                Here [x] represents multiplication by ${10^x}$.}
        \label{orbe_btoin}
        \begin{ruledtabular}
        \begin{tabular}{ldddddr}
              {Orbital}
              & \multicolumn{1}{c}{\textrm{GTO}}     &
                \multicolumn{1}{c}{\textrm{GRASP2K}}          &
                 \multicolumn{1}{c}{\text{$\Delta\epsilon_{\rm DC}$}} &
                 \multicolumn{1}{c}{\text{$\Delta\epsilon_{\rm Br}$}} &     
                 \multicolumn{1}{c}{\text{$\Delta\epsilon_{\rm Ue}$}} &     
                 \multicolumn{1}{c}{SE} \\     
              \hline 
              \multicolumn{7}{c}{B$^+$}\\
            \hline
            $1s_{1/2}$ & -8.18820  & -8.18820  & -1.078[-6]  & 1.205[-3] &  -1.348[-5] &  \\
            $2s_{1/2}$ & -0.87408  & -0.87408  &  3.724[-8]  & 4.683[-5] &   5.942[-7] &  \\ \\

              \multicolumn{7}{c}{Al$^+$}\\
              \hline
            $1s_{1/2}$ & -58.94477 & -58.94478 &  8.394[-6]  & 2.723[-2] &  -7.343[-4] &  1.350[-2]  \\
            $2s_{1/2}$ & -5.23616  & -5.23616  &  4.048[-6]  & 9.664[-4] &  -4.822[-5] &  9.440[-4]  \\
            $2p_{1/2}$ & -3.53257  & -3.53258  &  4.032[-6]  & 1.723[-3] &   1.136[-5] & -2.500[-5] \\
            $2p_{3/2}$ & -3.51519  & -3.51520  &  4.458[-6]  & 7.349[-4] &   1.136[-5] &  2.000[-5] \\
            $3s_{1/2}$ & -0.65347  & -0.65347  &  1.089[-7]  & 6.139[-5] &  -2.642[-6] &  5.900[-5]  \\ \\

              \multicolumn{7}{c}{Ga$^+$}\\
              \hline
            $1s_{1/2}$ & -384.21919 & -384.21918 & -1.465[-5]  & 4.854[-1] &  -2.484[-2] &  2.759[-1] \\
            $2s_{1/2}$ & -49.60651  & -49.60652  &  1.478[-5]  & 3.530[-2] &  -2.449[-3] &  2.908[-2] \\
            $2p_{1/2}$ & -43.74307  & -43.74302  & -5.118[-5]  & 6.234[-2] &   1.936[-4] & -7.880[-4] \\
            $2p_{3/2}$ & -42.71436  & -42.71431  & -4.938[-5]  & 4.065[-2] &   2.107[-4] &  1.499[-3] \\
            $3s_{1/2}$ & -6.88488   & -6.88493   &  5.125[-5]  & 3.734[-3] &  -3.663[-4] &  4.324[-3] \\
            $3p_{1/2}$ & -4.92169   & -4.92172   &  3.056[-5]  & 6.753[-3] &   3.833[-5] & -6.900[-5] \\
            $3p_{3/2}$ & -4.77952   & -4.77956   &  3.626[-5]  & 3.637[-3] &   4.111[-5] &  2.000[-4] \\
            $3d_{3/2}$ & -1.47943   & -1.47940   & -3.519[-5]  & 3.032[-4] &   3.061[-5] & -1.000[-5] \\
            $3d_{5/2}$ & -1.45974   & -1.45972   & -2.084[-5]  & 8.746[-4] &   3.038[-5] &  1.100[-5] \\
            $4s_{1/2}$ & -0.69963   & -0.69963   & -3.644[-8]  & 2.315[-4] &  -2.099[-5] &  2.700[-4] \\ \\

              \multicolumn{7}{c}{In$^+$}\\
              \hline
            $1s_{1/2}$ & -1033.04303 & -1033.04354 &  5.114[-4]  & 2.158     &  -1.166[-1] &  1.312 \\
            $2s_{1/2}$ & -158.20733  & -158.20736  &  3.044[-5]  & 1.956[-1] &  -1.871[-2] &  1.606[-1] \\
            $2p_{1/2}$ & -147.10243  & -147.10239  & -3.817[-5]  & 3.362[-1] &   6.087[-4] & -9.080[-4] \\
            $2p_{3/2}$ & -139.31685  & -139.31682  & -3.620[-5]  & 2.276[-1] &   1.086[-3] &  1.316[-2] \\
            $3s_{1/2}$ & -31.67750   & -31.67749   & -6.513[-6]  & 2.699[-2] &  -3.591[-3] &  3.156[-2] \\
            $3p_{1/2}$ & -27.15033   & -27.15031   & -2.113[-5]  & 5.005[-2] &   1.768[-4] &  2.770[-4] \\
            $3p_{3/2}$ & -25.70130   & -25.70128   & -1.989[-5]  & 3.007[-2] &   2.800[-4] &  2.527[-3]\\
            $3d_{3/2}$ & -17.78575   & -17.78573   & -2.199[-5]  & 1.392[-2] &   2.638[-4] & -2.080[-4] \\
            $3d_{5/2}$ & -17.49281   & -17.49280   & -3.223[-6]  & 5.087[-3] &   2.589[-4] &  2.580[-4] \\
            $4s_{1/2}$ & -5.58097    & -5.58098    &  3.750[-6]  & 3.905[-3] &  -6.601[-4] &  5.837[-3] \\
            $4p_{1/2}$ & -4.02420    & -4.02420    &  5.036[-7]  & 7.208[-3] &   6.176[-5] &  5.400[-5] \\
            $4p_{3/2}$ & -3.76860    & -3.76860    & -2.824[-6]  & 3.599[-3] &   8.024[-5] &  4.160[-4] \\
            $4d_{3/2}$ & -1.30374    & -1.30375    &  8.368[-6]  & 2.132[-4] &   6.288[-5] & -2.600[-5]  \\
            $4d_{5/2}$ & -1.26861    & -1.26861    &  3.999[-6]  & 9.058[-4] &   6.182[-5] &  3.200[-5] \\
            $5s_{1/2}$ & -0.63575    & -0.63575    & -4.827[-7]  & 3.259[-4] &  -5.627[-5] &  5.100[-4] \\
        \end{tabular}
        \end{ruledtabular}
\end{table*}


\subsection{B$^+$ and Al$^+$}

 The orbital energies from GTO for core orbitals, given in 
Table \ref{orbe_btoin}, are in excellent agreement with the numerical
data for both the ions. The largest difference between the two results is 
$8.394\times10^{-6}$ hartree, in the case of $1s_{1/2}$ orbital of Al$^+$. 
For remaining orbitals, of both the ions, the difference is even smaller. 
For $d$ virtual orbitals in Al$^+$, provided in Table \ref{virte_btoin}, 
the difference is of the order of $10^{-6}$ hartree or lower for the orbitals 
with principal quantum number $n\leq5$. For orbitals with $n>5$ the difference 
increases but still very small, the largest difference is 
$1.975 \times 10^{-4}$ in the case of $8d$. 

 Incorporating the Breit interaction, the change in the orbital energies, 
$\Delta \epsilon_{\rm Br}$ is positive for all the orbitals in both the ions.
This indicates relaxation of orbitals and a similar trend was observed in 
our work on group-IIB elements \cite{chattopadhyay-15}. As to be expected the inner
core core orbitals, which are closer to the nucleus, have larger corrections.
Quantitatively, $\Delta \epsilon_{\rm Br}$, $1.205\times10^{-3}$ hartree, 
for $1s_{1/2}$ orbital is two orders of magnitude higher than that of 
$4.683\times10^{-5}$ for the $2s_{1/2}$ orbital in B$^+$. Similarly, in 
Al$^+$, $\Delta \epsilon_{\rm Br}$, $2.723 \times 10^{-2}$ hartree, for 
$1s_{1/2}$ is three orders of magnitude larger than that of 
$6.139 \times 10^{-5}$ hartree for $3s_{1/2}$.

 The correction due to the Uehling potential, $\Delta \epsilon_{\rm Ue}$, is on 
average two orders of magnitude smaller than the Breit interaction for all 
the orbitals in both the ions. The other important difference from Breit is 
that, except for the $2s_{1/2}$ orbital in B$^+$, all the $s_{1/2}$ orbitals 
tend to contract. The remaining orbitals, $2p_{1/2}$ and $2p_{3/2}$, tend to 
relax after the inclusion of the Uehling potential. In terms of actual
contribution, similar to the trend of Breit interaction, 
$\Delta \epsilon_{\rm Ue (1s_{1/2})}$, $1.348\times10^{-5}$ hartree, is two 
orders of magnitude larger than $\Delta \epsilon_{\rm Ue (2s_{1/2})}$, 
$5.942\times10^{-7}$ hartree in B$^+$. Similarly, 
$\Delta \epsilon_{\rm Ue (1s_{1/2})}$ in Al$^+$ is two orders of magnitude 
larger than $\Delta \epsilon_{\rm Ue (4s_{1/2})}$. This trend is to be 
expected, as the vacuum polarization is attractive and short-range potential,
it has large effects on the orbitals with finite probability density within 
the nucleus. 

 For the self-energy correction, it is negligibly small in the case of B$^+$. 
So, we provide data for Al$^+$ only. As we observe from the table, like Breit 
interaction and vacuum polarization corrections, the self-energy correction 
${\rm SE}$ also is largest for the $1s_{1/2}$ orbital, and decreases with
increasing principal quantum number. Quantitatively, ${\rm SE}$ for 
$1s_{1/2}$ is $\approx 228$ times larger than that for $4s_{1/2}$. This is to 
be expected, as the inner core orbitals, which are closer to the nucleus 
have larger relativistic effects than the others. The other important 
observation is that, except the $2p_{1/2}$ orbital, ${\rm SE}$ is positive 
for all the orbitals.

\begin{table*}
        \caption{The orbital energies (in Hartree) from GTO compared with that
                from the GRASP2K\cite{jonsson-13} results for Tl$^+$. We also provide
                the contributions from the Breit interaction, vacuum polarization
                and the self-energy corrections. The self-energy corrections are 
                calculated using the code QEDMOD by Shabaev {et al.} \cite{shabaev-15}. 
                Here [x] represents multiplication by ${10^x}$.}
        \label{orbe_tl}
        \begin{ruledtabular}
        \begin{tabular}{ldddddr}
              {Orbital}
              & \multicolumn{1}{c}{\textrm{GTO}}     &
                \multicolumn{1}{c}{\textrm{GRASP2K}}          &
                 \multicolumn{1}{c}{\text{$\Delta\epsilon_{\rm DC}$}} &
                 \multicolumn{1}{c}{\text{$\Delta\epsilon_{\rm Br}$}} &     
                 \multicolumn{1}{c}{\text{$\Delta\epsilon_{\rm Ue}$}} &     
                 \multicolumn{1}{c}{SE} \\     
            \hline
	  $1s_{1/2}$ & -3164.43045 & -3164.43029 & -1.619[-4]  & 11.438    &  -1.658     &  7.801 \\
	  $2s_{1/2}$ & -569.10850  & -569.10847  & -2.862[-5]  & 1.289     &  -2.369[-1] &  1.190 \\
	  $2p_{1/2}$ & -545.21645  & -545.21644  & -1.574[-5]  & 2.168     &  -1.613[-2] &  1.012[-1] \\
	  $2p_{3/2}$ & -469.18398  & -469.18397  & -1.088[-5]  & 1.362     &   7.960[-3] &  1.468[-1] \\
	  $3s_{1/2}$ & -138.62965  & -138.62966  &  7.600[-6]  & 2.397[-1] &  -5.398[-2] &  2.766[-1] \\
	  $3p_{1/2}$ & -127.91899  & -127.91900  &  1.550[-6]  & 4.140[-1] &  -3.956[-3] &  3.027[-2] \\
	  $3p_{3/2}$ & -110.79473  & -110.79474  &  1.850[-6]  & 2.461[-1] &   2.336[-3] &  3.545[-2] \\
	  $3d_{3/2}$ & -93.35056   & -93.35059   &  2.126[-5]  & 1.802[-1] &   2.602[-3] & -2.150[-3] \\
	  $3d_{5/2}$ & -89.72674   & -89.72677   &  2.463[-5]  & 1.163[-1] &   2.462[-3] &  4.477[-3] \\
	  $4s_{1/2}$ & -32.55930   & -32.55932   &  1.562[-5]  & 4.945[-2] &  -1.363[-2] &  7.001[-2] \\
	  $4p_{1/2}$ & -27.91070   & -27.91071   &  1.151[-5]  & 8.834[-2] &  -8.320[-4] &  7.596[-3] \\
	  $4p_{3/2}$ & -23.69397   & -23.69398   &  1.113[-5]  & 4.601[-2] &   7.719[-4] &  8.733[-3] \\
	  $4d_{3/2}$ & -16.11031   & -16.11033   &  2.634[-5]  & 2.455[-2] &   7.758[-4] & -5.640[-4] \\
	  $4d_{5/2}$ & -15.31327   & -15.31328   &  2.968[-6]  & 1.038[-2] &   7.413[-4] &  1.169[-3] \\
	  $4f_{5/2}$ & -5.45750    & -5.45748    & -1.622[-5]  &-5.867[-3] &   5.503[-4] &  0.000  \\
	  $4f_{7/2}$ & -5.28151    & -5.28149    & -2.175[-5]  &-1.193[-2] &   5.396[-4] &  0.000  \\
	  $5s_{1/2}$ & -5.88606    & -5.88606    & -4.033[-6]  & 7.680[-3] &  -2.652[-3] &  1.377[-2] \\
	  $5p_{1/2}$ & -4.25160    & -4.25159    & -4.585[-6]  & 1.333[-2] &  -1.795[-5] &  1.330[-3] \\
	  $5p_{3/2}$ & -3.48442    & -3.48442    & -1.251[-6]  & 5.378[-3] &   2.810[-4] &  1.459[-3] \\
	  $5d_{3/2}$ & -1.16120    & -1.16120    &  4.185[-6]  & 4.544[-4] &   2.194[-4] & -6.200[-5] \\
	  $5d_{5/2}$ & -1.07329    & -1.07330    &  2.670[-6]  &  -1.067[-3] & 2.070[-4] &  1.230[-4] \\
	  $6s_{1/2}$ & -0.68952    & -0.68952    & -1.365[-6]  & 6.872[-4] &  -3.000[-4] &  1.519[-3] \\
        \end{tabular}
        \end{ruledtabular}
\end{table*}

\begin{table}
        \caption{The orbital energies for virtual orbitals (in Hartree) from GTO 
                 is compared with the GRASP2K\cite{jonsson-13} results for Al$^+$ 
                 and Ga$^+$. Here [x] represents multiplication by ${10^x}$.}
        \label{virte_btoin}
        \begin{ruledtabular}
        \begin{tabular}{lddr}
              {Orbital}
              & \multicolumn{1}{c}{\textrm{GTO}}     &
                \multicolumn{1}{c}{\textrm{GRASP2K}}          &
                 \multicolumn{1}{c}{\text{$\Delta\epsilon$}} \\     
            \hline
              \multicolumn{4}{c}{Al$^+$}\\
            \hline
            $3d_{3/2}$ & -0.05791  & -0.05791  &  4.859[-9]   \\ 
            $3d_{5/2}$ & -0.05791  & -0.05791  &  4.851[-9]   \\ 
            $4d_{3/2}$ & -0.03254  & -0.03254  &  5.076[-7]   \\
            $4d_{5/2}$ & -0.03254  & -0.03254  &  5.076[-7]   \\             
            $5d_{3/2}$ & -0.02072  & -0.02072  &  4.973[-6]   \\
            $5d_{5/2}$ & -0.02072  & -0.02072  &  4.973[-6]   \\ 
            $6d_{3/2}$ & -0.01430  & -0.01433  &  2.706[-5]   \\ 
            $6d_{5/2}$ & -0.01430  & -0.01433  &  2.706[-5]   \\ 
            $7d_{3/2}$ & -0.01039  & -0.01049  &  9.621[-5]   \\ 
            $7d_{5/2}$ & -0.01039  & -0.01049  &  9.620[-5]   \\ 
            $8d_{3/2}$ & -0.00781  & -0.00800  &  1.975[-4]   \\ 
            $8d_{5/2}$ & -0.00781  & -0.00800  &  1.975[-4]   \\ \\

              \multicolumn{4}{c}{Ga$^+$}\\
            \hline
            $4f_{5/2}$ & -0.03126   & -0.03087  &  -3.936[-4]   \\ 
            $4f_{7/2}$ & -0.03126   & -0.03087  &  -3.947[-4]   \\ 
            $5f_{5/2}$ & -0.02001   & -0.01946  &  -5.518[-4]   \\ 
            $5f_{7/2}$ & -0.02001   & -0.02024  &   2.304[-4]   \\ 
            $6f_{5/2}$ & -0.01388   & -0.01484  &   9.610[-4]   \\ 
            $6f_{7/2}$ & -0.01388   & -0.01406  &   1.798[-4]  \\ 
            $7f_{5/2}$ & -0.01015   & -0.00998  &  -1.695[-4]   \\ 
            $7f_{7/2}$ & -0.01015   & -0.00998  &  -1.672[-4]   \\ 
            $8f_{5/2}$ & -0.00762   & -0.00793  &   3.140[-4]   \\ 
            $8f_{7/2}$ & -0.00762   & -0.00793  &  -3.164[-4]   \\ 
            $9f_{5/2}$ & -0.00584   & -0.00624  &  -3.982[-4]   \\ 
            $9f_{7/2}$ & -0.00584   & -0.00622  &  -3.856[-4]   \\ 
            $10f_{5/2}$ & -0.00484  & -0.00506  &   2.148[-4]   \\ 
            $10f_{7/2}$ & -0.00484  & -0.00507  &   2.227[-4]   \\ 
            $11f_{5/2}$ & -0.00377  & -0.00410  &   3.231[-4]   \\ 
            $11f_{7/2}$ & -0.00377  & -0.00411  &   3.366[-4]   \\ 
            $12f_{5/2}$ & -0.01273  & -0.00351  &  -9.222[-3]   \\ 
            $12f_{7/2}$ & -0.01273  & -0.00350  &  -9.236[-3]   \\ \\

        \end{tabular}
        \end{ruledtabular}
\end{table}

\begin{table}
        \caption{The orbital energies for virtual {\em f} orbitals (in Hartree) 
                 from GTO are compared with the GRASP2K\cite{jonsson-13} results 
                 for In$^+$ and Tl$^+$. Here [x] represents multiplication by ${10^x}$.}
        \label{virte_tl}
        \begin{ruledtabular}
        \begin{tabular}{lddr}
              {Orbital}
              & \multicolumn{1}{c}{\textrm{GTO}}     &
                \multicolumn{1}{c}{\textrm{GRASP2K}}          &
                 \multicolumn{1}{c}{\text{$\Delta\epsilon$}} \\     
            \hline
              \multicolumn{4}{c}{In$^+$}\\
            \hline
            $4f_{5/2}$ & -0.03127  & -0.03127  &  -5.307[-8]   \\
            $4f_{7/2}$ & -0.03127  & -0.03127  &  -6.643[-8]   \\
            $5f_{5/2}$ & -0.02002  & -0.01985  &  -1.720[-4]   \\
            $5f_{7/2}$ & -0.02002  & -0.01978  &  -2.334[-4]   \\
            $6f_{5/2}$ & -0.01390  & -0.01407  &   1.736[-4]   \\
            $6f_{7/2}$ & -0.01390  & -0.01414  &   2.348[-4]   \\
            $7f_{5/2}$ & -0.01021  & -0.01379  &   3.585[-3]   \\
            $7f_{7/2}$ & -0.01021  & -0.01008  &  -1.211[-4]   \\
            $8f_{5/2}$ & -0.07786  & -0.07929  &   1.436[-4]   \\
            $8f_{7/2}$ & -0.00779  & -0.00795  &   1.621[-4]   \\
            $9f_{5/2}$ & -0.00609  & -0.00524  &  -8.482[-4]   \\
            $9f_{7/2}$ & -0.00609  & -0.00543  &  -6.567[-4]   \\
            $10f_{5/2}$ & -0.00490  & -0.00594  &  1.045[-3]   \\
            $10f_{7/2}$ & -0.00490  & -0.00575  &  8.538[-4]   \\ \\ 

              \multicolumn{4}{c}{Tl$^+$}\\
            \hline
            $5f_{5/2}$ & -0.03127  & -0.02002  &  -1.125[-2]   \\ 
            $5f_{7/2}$ & -0.03127  & -0.02002  &  -1.125[-2]   \\ 
            $6f_{5/2}$ & -0.02002  & -0.01390  &  -6.116[-3]   \\ 
            $6f_{7/2}$ & -0.02002  & -0.01390  &  -6.115[-3]   \\ 
            $7f_{5/2}$ & -0.01390  & -0.01021  &  -3.686[-3]   \\ 
            $7f_{7/2}$ & -0.01390  & -0.01021  &  -3.686[-3]   \\ 
            $8f_{5/2}$ & -0.01020  & -0.00782  &  -2.382[-3]   \\ 
            $8f_{7/2}$ & -0.01020  & -0.00782  &  -2.382[-3]   \\ 
            $9f_{5/2}$ & -0.00776  & -0.00618  &  -1.579[-3]   \\ 
            $9f_{7/2}$ & -0.00776  & -0.00617  &  -1.579[-3]   \\ 
            $10f_{5/2}$ & -0.00595 & -0.00500  &  -9.461[-4]   \\ 
            $10f_{7/2}$ & -0.00595 & -0.00500  &  -9.461[-4]   \\ 
            $11f_{5/2}$ & -0.00449 & -0.00413  &  -3.565[-4]   \\ 
            $11f_{7/2}$ & -0.00449 & -0.00413  &  -3.565[-4]   \\ 
            $12f_{5/2}$ & -0.00336 & -0.00347  &   1.168[-4]   \\             
            $12f_{7/2}$ & -0.00336 & -0.00347  &   1.165[-4]   \\ 
            $13f_{5/2}$ & -0.00297 & -0.00296  &  -1.145[-5]   \\ 
            $13f_{7/2}$ & -0.00297 & -0.00296  &  -1.203[-5]   \\ 
            $14f_{5/2}$ & -0.01735 & -0.00255  &   1.991[-2]   \\ 
            $14f_{7/2}$ & -0.01735 & -0.00255  &   1.990[-2]   \\ 
        \end{tabular}
        \end{ruledtabular}
\end{table}


\subsection{Ga$^+$ and In$^+$}

Like in the case of  B$^+$ and Al$^+$ the energies of both the core and 
virtual orbitals from GTO are in excellent agreement with the GRASP2K data. 
Quantitatively, for Ga$^+$, the energy of core and virtual orbitals agree 
up to $\approx 10^{-5}$ and $\approx 10^{-4}$ hartree, respectively. For 
In$^+$, the largest difference is of the order of $10^{-4}$ hartree, for 
the $1s_{1/2}$ orbital. 

For the corrections from the Breit interaction and vacuum polarization 
to the orbital energies, they follow the same trend as in B$^+$ and Al$^+$. 
That is, $\Delta\epsilon_{\rm Br}$ is positive for all the orbitals and 
$\Delta\epsilon_{\rm Ue}$ is negative for $s_{1/2}$ orbitals only. The 
magnitude of the corrections are, however, orders of magnitudes larger than 
that of Al$^+$. This trend is to be expected as Ga$^+$ and In$^+$ have 
stronger nuclear potentials than Al$^+$. In addition, like B$^+$ and Al$^+$,  
the inner core orbitals have larger corrections 
than the outer core orbitals. 

  The self-energy corrections, like in Al$^+$, ${\rm SE}$ is positive 
for all $s_{1/2}$ and $p_{3/2}$ orbitals and negative for all $p_{1/2}$ 
orbitals for both the ions. Among the remaining orbitals, all $d_{3/2}$ 
orbitals have negative and all $d_{5/2}$ orbitals have positive contributions. 
In terms of the magnitude of the corrections, it is one and two orders larger 
in Ga$^+$ and In$^+$, respectively than Al$^+$.


\subsection{Tl$^+$}

   Like in the other ions, the orbital energies are in excellent agreement
with the GRASP2K data. The largest difference of $1.619\times10^{-4}$ hartree
is observed in the case of $1s_{1/2}$ orbital. For the $f$ virtual 
orbitals, the difference is of the order of $10^{-2}$ hartree for 
$5f_j$ ($j=5/2, 7/2$), and smaller for the orbitals with $n>5$. The correction
from the Breit interaction $\Delta\epsilon_{\rm Br}$, as to be expected, is 
much more larger than the previous ions. Comparing with In$^+$, it is 
$\approx 5$, $7$, $9$, $12$ and $23$ times larger in magnitude for $1s_{1/2}$, 
$2s_{1/2}$, $3s_{1/2}$, $4s_{1/2}$ and $5s_{1/2}$ orbitals, respectively.
Roughly the same trend of contribution is also observed for all $p_{1/2}$ and
$p_{3/2}$ orbitals. For $3d_j$ and $4d_j$ ($j=3/2, 5/2$) orbitals 
$\Delta\epsilon_{\rm Br}$ is $\approx 13$ and $115$  times larger than
In$^+$. Among the other orbitals, $5p_{1/2}(5p_{3/2})$ and $5d_{3/2}(5d_{5/2})$ 
has the contributions of $1.333\times10^{-2}(5.378\times10^{-3})$ and
$4.544\times10^{-4}(1.067\times10^{-3})$ hartrees, respectively.

 Like the Breit interaction, corrections from the Uehling 
potential are also larger than the In$^+$ for all the orbitals. For instance,
$\Delta\epsilon_{\rm Ue}$ is  $\approx 14$, $12$, $15$, $20$ and $47$ times 
larger in magnitude for $1s_{1/2}$, $2s_{1/2}$, $3s_{1/2}$, $4s_{1/2}$ and 
$5s_{1/2}$ orbitals, respectively. It is $\approx 26$, $22$ and $13$ times 
for $2p_{1/2}$, $3p_{1/2}$ and $4p_{1/2}$ orbitals, respectively. For 
$p_{3/2}$ orbitals, the contributions are slightly smaller, $\approx 6$, $8$ 
and $9$ times for $2p_{3/2}$, $3p_{3/2}$ and $4p_{3/2}$ orbitals, respectively.
Looking into the corrections to $d_{3/2}$ and $d_{5/2}$, they both have the same
trend of contributions where $\Delta\epsilon_{\rm Ue}$ is $\approx 10$ and 
$12$ times larger than In$^+$ for $3d_{j}$ and $4d_{j}$ ($j=3/2, 5/2$), 
respectively. Among the other orbitals, $4f_{j}$ have the same order of 
contributions as $4d_{j}$ orbitals.

Considering the correction from the self-energy, there is an important
difference in the sign of ${\rm SE}$ in comparison to the previous ions. The
${\rm SE}$ is positive for all the $p_{1/2}$ orbitals, which is in opposite
phase to the previous ions. Comparing with In$^+$, in terms of magnitude, 
among the $s_{1/2}$ orbitals, $5s_{1/2}$ and $1s_{1/2}$ has the highest and 
lowest contributions of $\approx27$ and $6$ times, respectively.
Similarly, among the $p_{1/2}$($p_{3/2}$) orbitals, the largest and smallest
change from In$^+$ are $\approx 111$($\approx 20$) and $109$($11$) 
times for $4p_{1/2}$($4p_{3/2}$) and $3p_{1/2}$($2p_{3/2}$) orbitals, 
respectively. For ${\rm SE}$ in the case of $d_{3/2}$ and $d_{5/2}$, there is 
a large change for $d_{5/2}$ than $d_{3/2}$. 
Quantitatively, it is $\approx 17$ and $36$ times, respectively for $3d_{5/2}$ 
and $4d_{5/2}$ in comparison to $\approx 10$ and $21$ times, respectively for 
$3d_{3/2}$ and $4d_{3/2}$ orbitals.


\bibliography{pol_group3a}
\bibliographystyle{apsrev4-1}

\end{document}